Title: An Observational Study of a "Rosetta-Stone" Solar Eruption

Short Title: "Rosetta-Stone" Eruption

E. I. Mason[1]
Heliophysics Science Division, NASA Goddard Space Flight Center, Greenbelt, MD 20771
emily.mason@nasa.gov

Spiro K. Antiochos
Goddard Space Flight Center, 8800 Greenbelt Rd., Greenbelt, MD 20706
spiro.antiochos@nasa.gov

Angelos Vourlidas
Johns Hopkins Applied Physics Laboratory, 11100 Johns Hopkins Road, Laurel, Maryland 20723
angelos.vourlidas@jhuapl.edu

---

[1] NASA Postdoctoral Program Fellow

## ABSTRACT

This Letter reports observations of an event that connects all major classes of solar eruptions: those that erupt fully into the heliosphere versus those that fail and are confined to the Sun, and those that eject new flux into the heliosphere, in the form of a flux rope, versus those that eject only new plasma in the form of a jet. The event originated in a filament channel overlying a circular polarity inversion line (PIL) and occurred on 2013-03-20 during the extended decay phase of the active region designated NOAA 12488/12501. The event was especially well-observed by multiple spacecraft and exhibited the well-studied null-point topology. We analyze all aspects of the eruption using SDO AIA and HMI, STEREO-A EUVI, and SOHO LASCO imagery. One section of the filament undergoes a classic failed eruption with cool plasma subsequently draining onto the section that did not erupt, but a complex structured CME/jet is clearly observed by SOHO LASCO C2 shortly after the failed filament eruption. We describe in detail the slow buildup to eruption, the lack of an obvious trigger, and the immediate reappearance of the filament after the event. The unique mixture of major eruption properties observed during this event places severe constraints on the structure of the filament channel field and, consequently, on the possible eruption mechanism.

Subject keywords:

## 1. INTRODUCTION

Solar eruptive events (SEE) have long been observed to have one of three distinct forms: a CME that ejects both plasma and new magnetic flux into the heliosphere (e.g., Linker et al. 2003), a jet that ejects plasma and magnetic wave energy but no closed flux (e.g., Raouafi et al. 2016), and a failed eruption that ejects neither plasma nor magnetic field (e.g., Schrijver 2009). Irrespective of the form, all SEEs are believed to be due to the explosive release of the free energy stored in the magnetic field of a filament channel (Parenti 2014, Georgoulis et al. 2019). The eventual form of the eruption, therefore, must be due to the nature of the filament field and the field and plasma of the surrounding corona; consequently, determining the mechanism for an eruption and the physical reason for its form requires measuring as accurately as possible the magnetic and plasma structure of the corona and filament channel (e.g. Patsourakos et al. 2020 and references therein). As we show below, the topology of the coronal magnetic can be determined accurately from high-resolution EUV images and can also be inferred from observations of the photospheric magnetic field, but accurate quantitative measurements of the coronal field are not yet available. Furthermore, determining the structure of the filament magnetic field from observations has long been one of the most important, unsolved problems in solar astronomy.

Two general models have been proposed for the filament field: a twisted flux rope (e.g., van Ballegooijen & Martens 1989) and a 3D sheared arcade (Antiochos et al. 1994). The key difference between the models is that the flux rope has a coherent global twist due to a net current running along the filament channel. As a result, the structure can be susceptible to an ideal instability such as a kink or torus (Fan 2005, Kliem & Török 2006, Liu 2008). The sheared arcade, on the other hand, has no global coherence and requires a resistive process – magnetic reconnection – for eruption (Antiochos et al. 1999). In principle, it should be possible to distinguish between the two models by analyzing the source region and detailed dynamics of an eruption; but the dynamics also depend critically on the structure of the surrounding magnetic field. Large, energetic and well-observed eruptions tend to occur in delta-spots (Antiochos 1998, Liu et al. 2005), active region complexes with multiple coronal flux systems; consequently, it is difficult to separate the eruption dynamics into those due to the nature of the filament channel versus the surrounding field interactions. As a result, observations of eruptions, to date, have not definitively validated or invalidated either filament channel model.

In this Letter we describe a unique event that involves all three forms of eruption and poses severe challenges for the filament channel/eruption models. The event occurs in, perhaps, the simplest possible coronal magnetic field, the so-called null-point topology. This is a key point, because it implies that any complexity in the eruption must be due to the structure of the filament field rather than to the surrounding corona.

Null-point topologies have been reported since the days of *Yohkoh* and the identification of the anemone jet (Shibata 1994). This category of coronal events encompasses the range from tiny jets embedded in coronal holes to large pseudostreamers that can span multiple prominences. While there is no overarching count of null-point topology occurrence frequency, campaigns to study jets have catalogued thousands, and pseudostreamers can be observed regularly on the limb during all but the quietest periods of the solar cycle. Most commonly, they are formed by the emergence or migration of a minority polarity region on the photosphere into a larger unipolar background of either open or closed flux (Moore et al. 2010, Paraschiv et al. 2010, Raouafi et al. 2016). This creates a closed-field region in the corona bounded by what is referred to interchangeably as a dome or fan surface, and an outer spine emanating from a null point on this dome and connected out into the open heliosphere or to some far distant closed field region (e.g., Pariat et al. 2009). Null-point topologies constitute a common and important set of coronal features, since their magnetic structures and locations, often adjacent to or embedded within open-field regions are the source of various dynamics that could transfer coronal energy and mass into the solar wind.

Previously (Mason et al. 2019), we reported on a specific subset of these structures, called raining null-point topologies (RNPTs), which are decaying active regions with null points located between 50-150 Mm in height that form on open/closed boundaries. The key properties of RNPTs are that they are relatively large and either in or close to a coronal hole boundary. As a result, they are easily observed. Here we discuss the end-of-life period of Event 3 from Mason et al. 2019, an RNPT that erupts in a manner not previously observed, including a narrow coronal mass ejection (CME) resulting from a particularly intriguing prominence eruption.

The RNPT consists of highly-decayed active region with a well-established circular prominence that erupts to produce a narrow, poorly-structured jet-like CME without fully erupting the prominence material, and without any discernible flaring behind the eruption. The eastern half of the prominence simply rises slowly for over 30 hours without any evidence of energization, then accelerates to over 350 km s$^{-1}$ briefly before decelerating and draining the cool plasma onto the non-erupted half of the preexisting prominence. As we shall discuss in greater detail in subsequent sections, many of the characteristics of this event are unusual when compared to traditional eruptive models. Each of these characteristics – CME, jet, and failed eruption – has been seen in many other events, but the presence of all three is unique. Furthermore, the event poses severe challenges to the eruption models, both the ideal and the reconnection-driven.

## 2. 2016 MARCH 13 ERUPTION

*2.1 Global Magnetic Structure and Early Dynamics*

The source of the event began as an active region (AR) designated NOAA 12488, first identified on 2016 January 20 (Figures 1ab show NOAA 12488 6 days later). The AR transited the solar disk and was labeled NOAA 12501 when it reappeared on the east limb on February 13. The initial structure of the AR was a multipolar cluster with a magnetic class going up to βγ, but it had settled into an almost uniform α designation during its second solar rotation (Figure 1cd).

As early as January 25, cool dark material can be seen collecting in H-alpha observations (Figure 1b). This prominence grows around the polarity inversion line (PIL, Figure 1d) until it entirely encompasses the PIL on March 1 in the Extreme Ultraviolet Imager (EUVI; Wuelser et al. 2004) imagery (Figure 1ef) from the Solar-Terrestrial Relations Observatory (STEREO; Kaiser 2005) Ahead spacecraft. This prominence has several partial eruptions, visible in the Solar Dynamics Observatory Atmospheric Imaging Assembly (SDO AIA; Lemen et al. 2011) field of view on February 18 and in the EUVI-A field of view on March 1, March 4, and March 10. STEREO was located 163° east of the Sun-Earth line during the far side observations of early March. None of the partial eruptions significantly altered the shape, size, or amount of cool material in the prominence.

Prominence eruptions are often classified into two broad categories: the dramatic but relatively low-energy liftoffs associated with the vast, long-lived quiet Sun prominences that encircle significant portions of the solar circumference, and the much smaller and shorter-term eruptions occurring from within ARs, often accompanied by flares (Mackay et al. 2010). Both types can be home to partial eruptions, with large proportions of the prominence's cool plasma and overall magnetic structure left after some escapes (Gibson & Fan 2006, Tripathi et al. 2013, Parenti 2014). Debate also continues about whether there is one or more flux ropes involved in the partial eruption scenario, and whether they remain intact during the ejection (Liu et al. 2012, Zheng et al. 2019, Joshi et al. 2020a, 2020b). This event's importance is due in part because it developed within an AR, but due to the highly-decayed state of the AR at the time of eruption, the prominence's height and lifespan more closely reflect a quiet-Sun prominence. The early eruptions are similar to typical AR prominence eruptions, but the eruption of 13 March discussed below reveals characteristics of both types.

During its second visible rotation, the AR had decayed sufficiently to form an RNPT, which was reported in Mason et al. (2019). Context imagery of the RNPT on the east and west limbs can be seen in Figure 2a and 2b, respectively. The main concentration of magnetic field is centered near the southern border of the polarity inversion line, and this area continued to flare intermittently during February, with occasional attendant arcades, in STEREO-A EUVI imagery. The structure hosted tens of hours of coronal rain along the spine and down the fan surface, which had mostly stopped once it reached the west limb on February 27. There were still a few condensations forming on the northern border of the dome structure as it reappeared on the east limb on March 11.

The PFSS extrapolations of the structure's on-disk transits in January and February indicate that the decaying active region is situated under a helmet streamer (Figure 2d). However, LASCO C2 coronagraph images show a bright pseudostreamer-like spine extending from the vicinity of the RNPT, which is in an otherwise isolated region (Figure 2e). A key point here is that the PFSS does not take into account the extra magnetic stress due to the filament channel and, hence, would tend to underestimate the amount of open flux. It is therefore an open question whether the spine of this structure is open or simply closed to a very distant point in the Sun's southern hemisphere; this problem will be discussed more fully in Section 3, as the dynamics of the eruption raise further questions about the overlying magnetic topology.

*2.2 Prominence Rise Phases*

The prominence under the fan surface began a slow rise phase at approximately 10:00 UT on March 12, which lasted until 19:51 UT on March 13. During this period, the average rise speed was approximately 0.2 km s$^{-1}$. There are several points, visible in the AIA composite movie (Figure 3a and Movie 1) made using the JHelioviewer application (Müller et al. 2017), wherein loops near the top of the prominence lift more rapidly than the bulk rise speed and then drain. This is also apparent in Figure 3b, where the top line of the prominence shows "notches", the clearest example found around 15:00 and indicated by a white arrow. Figure 3b covers the slow rise phase through the beginning of the eruption in AIA 304 Å, at a cadence of 150 seconds.

While the prominence does complete a closed ring under the fan surface, the distribution of material is not uniform; the northeastern-most portion of the ring is substantially depleted of cool plasma, as a result of a previous partial eruption from that location. That event occurred on March 10$^{th}$ and was visible only in

STEREO-Ahead data. Cool plasma evacuation is often seen in destabilizing prominences prior to CME releases (e.g., Vourlidas et al. 2012) and is one of the characteristics in which this eruption resembles classic CME dynamics.

The rapid rise phase begins around 19:30 UT on March 13 with a short period of fast acceleration until 19:57 UT (Figure 3c). The rapid rise then proceeds at a roughly constant velocity of 388 km s$^{-1}$ until 20:08 UT before decelerating. As can be seen in the AIA 304 Å 12-second cadence data in Movie 2 (Figure 3d), the eruption is partial rather than complete: the majority of the prominence material does not escape, but instead flows back down towards the PIL on the western half of the prominence. The hotter plasma populating the dome and spine, seen in the AIA 171 and 193 Å portions of Movie 2, expands outward and leaves the AIA field of view (FOV) by 20:02 UT; this is the material which makes the CME shown in Figure 4a, corroborated by the co-temporal data from the STEREO-A EUVI instruments in 195 and 284 Å.

## 2.3 Partial Eruption Dynamics

Several details of the partial eruption are worthy of deeper focus. The most central of these is the complete lack of flare brightening throughout the entire rising phase of the prominence. The eruption appears to occur without any distinct energizing trigger; there are no events listed on either the RHESSI or GOES flare databases anywhere in the vicinity, and there is similarly no sign of brightening near the footpoints or within the body of the prominence until well into the eruption, as visible in STEREO-A EUVI imagery (see Movie 3, Figure 4b).

Once the eruption is underway intense brightenings appear along both legs of the prominence. They are visible in both AIA and EUVI 304 Å beginning around 19:48 UT and may indicate the presence of current sheets. The reconnection can be seen along the northern leg in AIA, and along the southern leg in EUVI, almost exclusively. Along the southern leg in particular, the reconnection develops into a very well-defined line that lasts until 20:08 UT; as we will discuss more fully in Section 3, the high definition of this line hints at a magnetic border which the structure may abut, restricting the expansion of the prominence as it reconnects and relaxes.

Another noteworthy characteristic is the band of prominence material that extends beyond the edge of the FOV in both SDO AIA 304 and STEREO-A EUVI 304 Å observations beginning at 21:02 UT. Unlike the preponderance of prominence plasma which outlines clearly closed loops throughout the eruption, it is unclear whether this plasma was escaping on open field lines which have reconnected during the eruption or whether it was just on a longer loop whose apex is beyond the FOV. There is no sign of this particular band of material in H-alpha data from the GONG network (Figure 4c), while much of the closed-loop partial eruption is visible, so it is possible that that band was on open field and was being heated as it was driven into the upper corona.

As the majority of the cool plasma drained back towards the surface, beginning at 20:31 UT, it did not drain back onto the eastern half of the prominence from which it rose; instead, it drained onto the western half. The flows can be followed for some distance around the prominence during the draining phase, almost exactly to the location of the arcades which formed over the eastern half of the prominence in the 193 Å data. Since the reconnection referenced previously did not allow the plasma to escape into the upper corona, it is possible that the eastern and western portions of the prominence reconnected with each other. This would channel the cool plasma onto the opposite half of the prominence. This may be related to the small cylindrical brightening near the surface on the northern side of the prominence that appeared around 20:22 UT in AIA 304 Å and 5 minutes later in the hotter channels. The material that flowed down out of it clearly continued onto the western half of the prominence. This may be the means by which the cool plasma drained onto the side opposite that on which it originated.

*2.4 Resultant CME*

Several coronagraphs recorded a narrow CME emanating from this eruption. It first emerged into the lower FOV of LASCO C2 at 20:39 UT. Likely due to projection effects, the apparent border of the CME overlapped with the signature of the helmet streamer to the south, which appeared to be subsequently displaced southward slightly due to the CME. The CME itself measured only 16° in half-width and had an outward speed of 679 km s$^{-1}$ as measured by SOHO, and 727 km s$^{-1}$ as measured by STEREO-A; both measurements were determined using StereoCat.

The leading edge and core of the CME are both apparent in the C2 imagery, but the core is relatively dim and there is no defined cavity present – overall, the CME is poorly structured and doesn't appear to be carrying much dense material. This is consistent with the disrupted nature of the eruption in the EUV and the absence of coherent features leaving the EUV FOVs that would signify the eruption of a magnetic flux rope. As we discuss next, the partial failure and lack of coherence in the escaped structure point towards the magnetic topology over the erupting prominence (e.g,. Chintzoglou et al. 2017).

## 3. IMPLICATIONS FOR THEORY AND MODELS

*3.1 Jets vs. CMEs vs. failed eruptions*

We consider this eruption a "Rosetta Stone" because it connects all three forms of solar eruption: jets, CMEs, and failed eruptions. Like a jet, the eruption was narrow in angular span, rooted in a relatively simple magnetic bipole structure, and did not exhibit much internal structure (i.e., a classic 3-part CME). However, the speed of the ejected material and the leading edge's shape profile in white-light observations were more similar to a CME than a jet. More recent CME identification algorithms have broadened the working definition of an ejection to include much more narrow-angle events than previous manually-created event databases (Webb & Howard 2012).

Our observations imply a blurring of the distinctions between jets and CMEs. In principle, the physical differences are clear: jets eject only magnetic waves and plasma into the heliosphere whereas CMEs eject a large amount of closed magnetic flux as well. We find that the ratio of wave to closed flux ejection can vary greatly from event to event, and even within a single event. MHD simulations show that the same magnetic drivers are at work in both scenarios (i.e., Wyper et al. 2017), and observationally, events such as the 2016 March 13 eruption show that the same characteristics occur in the range that lies between the two types of eruptions. This event adds to the growing consensus that jets and CMEs are merely two ends of a magnetically-driven eruption continuum that spans coronal structures across a broad spectrum of size scales (Vourlidas et al. 2013, 2017; Kumar et al. 2021).

One important new feature of our event is that it included a failed eruption, which is neither a jet nor CME, but it is generally associated with so-called compact flares. The defining characteristic of all flares is intense heating of the coronal plasma, often most easily detected by the bright flare ribbons on the chromosphere (Schmieder et al. 2015 and references therein). Our event, however, is distinguished by the absence of observable flaring prior to the main phase of the eruption. Reconnection signatures do not appear until well into the eruption on March 13 (Figure 4b), when ribbons form on either side of semicircle of the erupted prominence, indicating that classic flare reconnection has occurred. It is possible that a slower and lower-energy form of reconnection occurs in older, weaker magnetic structures, which does not produce the same EUV enhancements that are observed in more traditional eruption events.

The event lacks traditional signatures of filament eruption as well: no sigmoid forms before the eruption, there is no apparent footpoint brightening anywhere around the PIL before or during either rise phase (slow or fast), and there is no darkening around the filament. Furthermore, when the prominence rises, there is very little apparent twist among the various visible flux tubes, certainly less than the typical critical values for a kink instability. While the appearance of the event on the limb allowed for excellent front and back views using SDO and STEREO, it utterly negated the ability to analyze the underlying photospheric field motions before and during the event. Therefore, it is impossible to tell if the event was preceded by emergence or cancellation around the PIL. Significant emergence seems unlikely, however, given that the system consisted of an AR that had been decaying for many days.

This event, while unusually well-positioned and informative, is not an exception: we have identified several other partial eruptions of ring-like prominences in null-point topologies which have some characteristics, both in EUV and white light observations, in common with the event of this paper (i.e., 2015 January 12 in NOAA AR 12261). These eruptions are similar in some ways to the pseudostreamer eruptions documented in Wang (2015); like those, the eruptions are confined in angle and are secondary to prominence eruptions. However, it is unclear whether the event reported here was situated in open field on both sides, and we conjecture that interchange reconnection plays a critical role in the dynamics of the failed eruption. We plan to investigate these similarities and differences in these events in greater detail, using broader multi-wavelength studies and simulations to probe the underlying magnetic dynamics. The advent of coronal magnetic field data from DKIST and high-resolution spectral data from Solar Orbiter should provide a greater depth of useful data from observations of these ubiquitous structures.

### 3.2 Constraints on the Eruption Mechanism

A central question for all solar eruptions is the nature of the magnetic topology of the filament channel and of the surrounding overlying coronal field. Given the HMI observations of simple parasitic polarity region and the presence of the ring-shaped polarity inversion line with cool filament material, the global basic topology is clearly that of the well-known embedded bipole (Antiochos 1990), but the exact nature of the surrounding field is unclear.

In principle, we should be able to determine the initial global topology by following the outer spine and seeing if it is open or closed. However, this is difficult to do for this case where the outer spine is faint and extends outward to great heights. As previously stated, PFSS extrapolations derived from SDO HMI data, spanning the disk crossings of NOAA 12488/12501, always present the dome structure as inside the closed field of a helmet streamer bordering the northern polar coronal hole. LASCO imagery at several points before and after the failed eruption shows a bright, compact spine extending out into the solar wind from the AR's vicinity, in which case the surrounding field would be open. This discrepancy may well be due to the fact that the PFSS model does not include the magnetic stress due to the filament channel, which would tend to push upward the surrounding field and open it. After the null-point topology has emerged on the eastern limb but shortly before the eruption, however, there is no outer spine visible in LASCO; the spine signature reappears only after the eruption takes place. Furthermore, the dome structure rises rapidly out of the FOV of SDO and STEREO-A early in the eruption, showing no deviation or deformation as one might expect if it was being hampered by the overlying closed field lines of a helmet streamer. Numerical simulations of null-point topologies inside closed field, but near a coronal hole boundary, have shown that breakout reconnection can readily allow the null to move from the closed to open field region (Edmondson et al. 2010, Masson et al. 2014, 2019, Wyper et al. 2021). Consequently, there may well be a minimal difference in the eruption dynamics between a pseudostreamer that is fully in the open field versus one that is very near to it, and the LASCO imagery may provide observational support for this dynamic.

An important result is that during the partial eruption the prominence loops are observed to expand much more in height than they do in width (Figure 4d), appearing confined on both the northern and southern borders. This is the typical behavior for coronal jets (e.g., Wyper et al. 2018), where the size of the closed field system is small compared to a solar radius and the free energy in the filament channel is small compared to the energy of the surrounding open field. As can be seen from Fig.1 and 2, the closed dome has a size ~ 0.1 $R_S$ and the photospheric field is considerably weaker than that of a strong active region at this point of the decay process (the strongest points of field measure about 300 G on 14 March, when the structure is sufficiently on-disk to observe with HMI). Even jets, however, frequently show complete eruption of cool filament material out along the outer spine (e.g. Sterling et al. 2015). Therefore, the failure of the filament eruption is unusual, especially since a CME did accompany the event, so this was clearly not a typical confined flare.

The combination of a successful CME with a failed eruption from the same filament channel, during a single event, and from the simple topology of an embedded bipole poses severe challenges to present models of solar eruptions. Our observations imply that the initial filament channel broke up into or consisted of three distinct layers in height. The topmost layer underwent a successful eruption. This layer did not exhibit observable cool material, but it must have contained a large magnetic shear and free energy, because it produced an observable CME as well as a jet. The middle layer contained considerable cool material and rose in height by, at least, an order of magnitude from its initial height above the chromosphere, but then its rise stalled, and it appeared that all the cool material drained back onto the chromosphere. Finally, there is a bottom layer evident in Fig. 4 that appeared to not participate at all in the eruption.

As discussed in the Introduction, two topologies have been proposed for a filament channel: a 3D sheared arcade and a twisted flux rope. If the latter, then either two or more distinct twisted flux ropes were present prior to the eruption or the initial rope broke up into three flux ropes as a result of two distinct flare reconnections. The observation that a filament channel eruption leaves a bottom section that contains part or even all of the cool material undisturbed is not unusual. Within the context of the ideal eruption models, such as the kink instability, this observation has been attributed to the initial twisted flux rope splitting into two pieces as a result of some lower section being held down by heavy chromospheric material and separating from an upper erupting section via flare reconnection (e.g., Gibson & Fan 2006). In order to apply this scenario to our event, the flux rope would have to split at two well-separated heights. It is very difficult to understand, however, why the split between the top and middle section would ever occur, given that the middle section is clearly not held down and rises up to substantial heights. Furthermore, there is no evidence for the flare reconnection that would accompany the split between the top and middle layers.

The conclusion from this argument is that if the filament channel has a flux rope topology, then it must contain at least two distinct ropes prior to eruption. Note that this conclusion is not new; some authors have even proposed triple-decker flux ropes stacked on top of each other (Joshi et al 2020); also, filament/prominences in which the cool material appears to lie in two layers are sometimes observed and have been attributed to stacked twisted flux ropes (e.g., Liu et al 2012 and references therein). The key difference between our event and that of previous authors is that we observed a single event, not a sequence of eruptions separated by hours, and our prominence shows no evidence for multiple layers. In fact, the whole system is very benign, consisting of an old, highly-decayed active region in the ubiquitous topology of an embedded bipole, with a standard circular polarity inversion line and filament. It is not at all clear how topologically distinct multiple flux ropes would form and survive for many days in such a system, and there is no evidence for the twisted field lines of a high-lying flux rope in the SDO EUV images.

If, on the other hand, the filament channel topology had a sheared arcade topology, then splitting up into three or even more layers can readily occur, because in the standard model there are no connections between the sheared flux at different heights (e.g., Antiochos et al 1994). Furthermore, maintaining the filament channel is not an issue, shear flows and/or helicity condensation (e.g., Knizhnik et al. 2019) could slowly build up the sheared field for many days without any need for flux emergence or cancellation. The problem for this model, however, is the observed "Rosetta Stone" dynamics that contains all three forms of eruption.

Irrespective of any model, it is clear from our observations that the filament channel does break up into three layers that undergo very different evolutions. By itself, the narrow CME eruption has been reproduced by numerous simulations (e.g., Masson et al. 2013). If there is sufficient flux in the filament channel with sufficiently strong shear, it can erupt outward as a flux rope that survives being fully consumed by interchange reconnection with the surrounding open field. Such an eruption will naturally leave some of the sheared flux behind, because it is energetically prohibitive to produce a flare current sheet that reaches all the way down to the photosphere. Conversely, a failed eruption by itself is also straightforward to reproduce. If the parasitic polarity occurs in a large, closed region, then any eruption from within the parasitic polarity dome is likely to be stopped by the overlying flux (Wyper & DeVore 2016). It is commonly observed that flares from filament channels at the centers of active regions are non-eruptive, presumably due to the presence of strong overlying flux (e.g., Yan et al. 2018).

For our event, however, the reason for the failed eruption presents a challenge. Overlying flux was clearly insufficient to prevent the narrow CME eruption. If anything, this CME should have weakened the overlying flux, making it easier for the mid-level filament flux to erupt as well. At the very least, the filament bearing flux should have produced a jet via breakout reconnection. It is intriguing that this flux contained the dense filament material, which suggests that perhaps the weight of the plasma played some role. On the other hand, most of the material drained back down, but there was no evidence that the filament flux subsequently continued to rise. The observations present no immediate explanation for the eruption failure.

One enticing interpretation for the eruption's failure relies on the chaotic nature of its position within the global magnetic environment, seated on the boundary between the northern polar coronal hole and a well-defined helmet streamer to the south. The open/closed boundary is often described as existing in a state of constant change, with magnetic flux being processed continuously as individual magnetic field lines undergo regular interchange reconnection (Antiochos et al. 2011, Viall & Borovsky 2020). However, when discussing specific topologies, this region is still often thought about as a hard border. It is possible instead that a structure such as this null-point topology in fact exists in a more complex reality, where some of the overlying field is open and some is closed at any given moment. The spine observed in LASCO could potentially be extending out between the loops of helmet streamer legs, and the overlying helmet streamer field lines could aid in the failure of the prominence to erupt fully. Conversely, the entire structure could have slipped back and forth under the helmet streamer boundary several times during its multi-rotation lifetime, explaining why the spine was clearly observed during some periods and absent in others. Furthermore, the failed eruption was confined latitudinally on both the northern and southern sides. If the open-field region on the northern border is diffuse and constantly shifting, it could make large-scale reconnection with the prominence less likely to occur, and decrease the likelihood of the material escaping in a successful eruption.

A key aspect of the event is that it should be relatively easy to model. The photospheric flux distribution is perhaps the simplest possible with modest field strengths. Furthermore, the system was well observed for many days. Although the photospheric field was not observed exactly during the event since it was on the East limb, the data are available a few days later. The only important feature that is truly unconstrained is the structure of the filament channel field. Consequently, it should be possible to test rigorously the various models for filament channels and eruption by inserting a sheared arcade and/or multiple flux ropes in an attempt to reproduce all the manifest properties of this event. As a result of such a study our event may very well turn out to be a "Rosetta Stone" for interpreting solar eruptions.

## 4. ACKNOWLEDGMENTS

EIM's research was supported by an appointment to the NASA Postdoctoral Program at the NASA Goddard Space Flight Center, administered by Universities Space Research Association under contract with NASA. SKA and EIM would like to acknowledge support from the NASA HSR and GI programs. AV was supported by NASA grant 80NSSC19K1261 and 80NSSC19K0052.

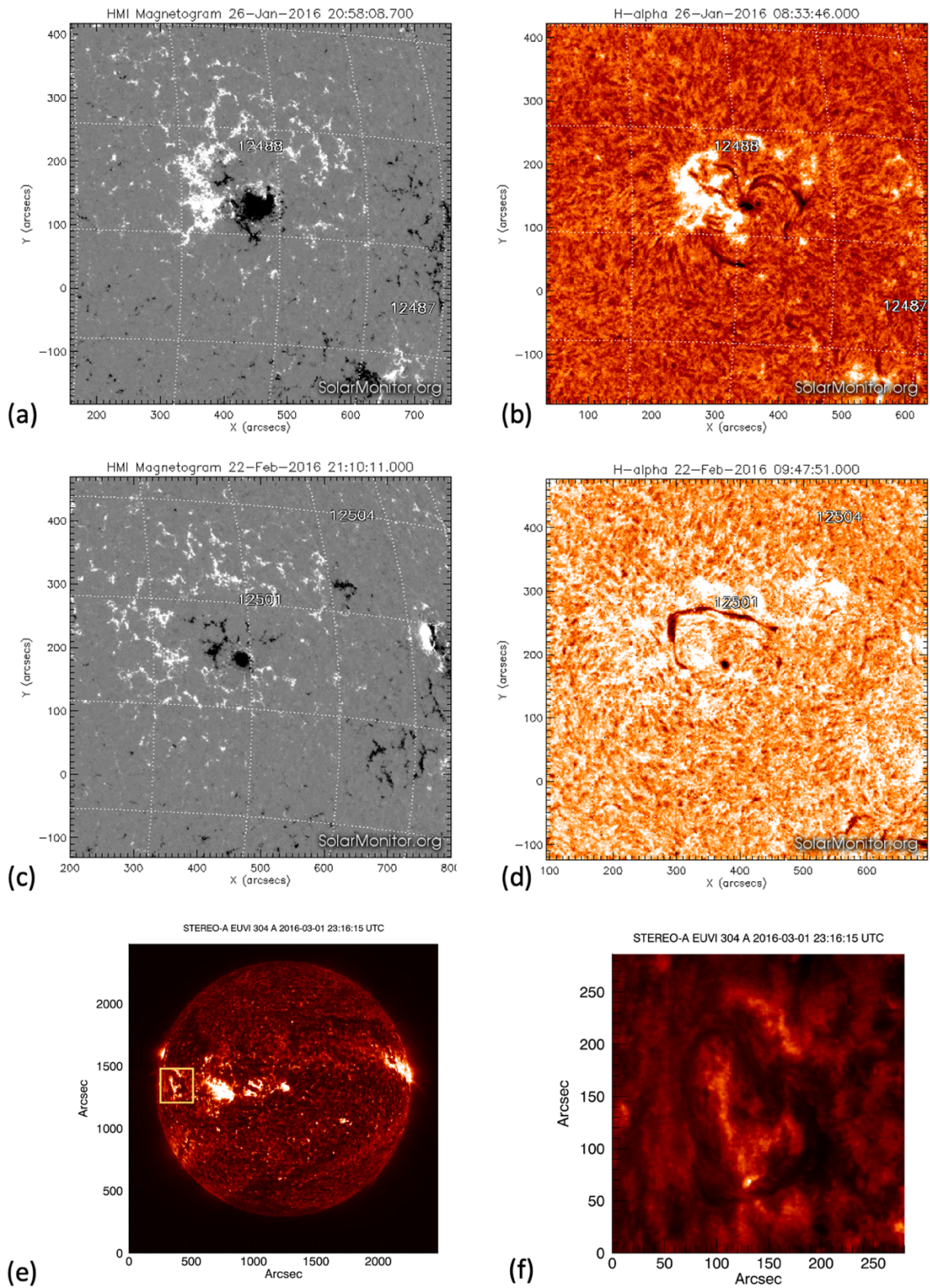

Figure 1: (a) SDO HMI magnetogram of NOAA AR 12488 on 2016 January 26. (b) Global H-alpha Network (GHN) data for the same AR, showing the early components of the prominence beginning to encompass the negative-polarity region seen in (a). (c) Analogous to (a), but for the following Carrington rotation when the AR is renamed to NOAA 12501. (d) Analogous to (b), showing how much the prominence has consolidated. (e) STEREO-A 304 Å image showing the AR just after clearing the eastern limb on 2016 March 1, at which point the prominence is a closed structure. (f) A zoomed-in view of (e) focused on the prominence.

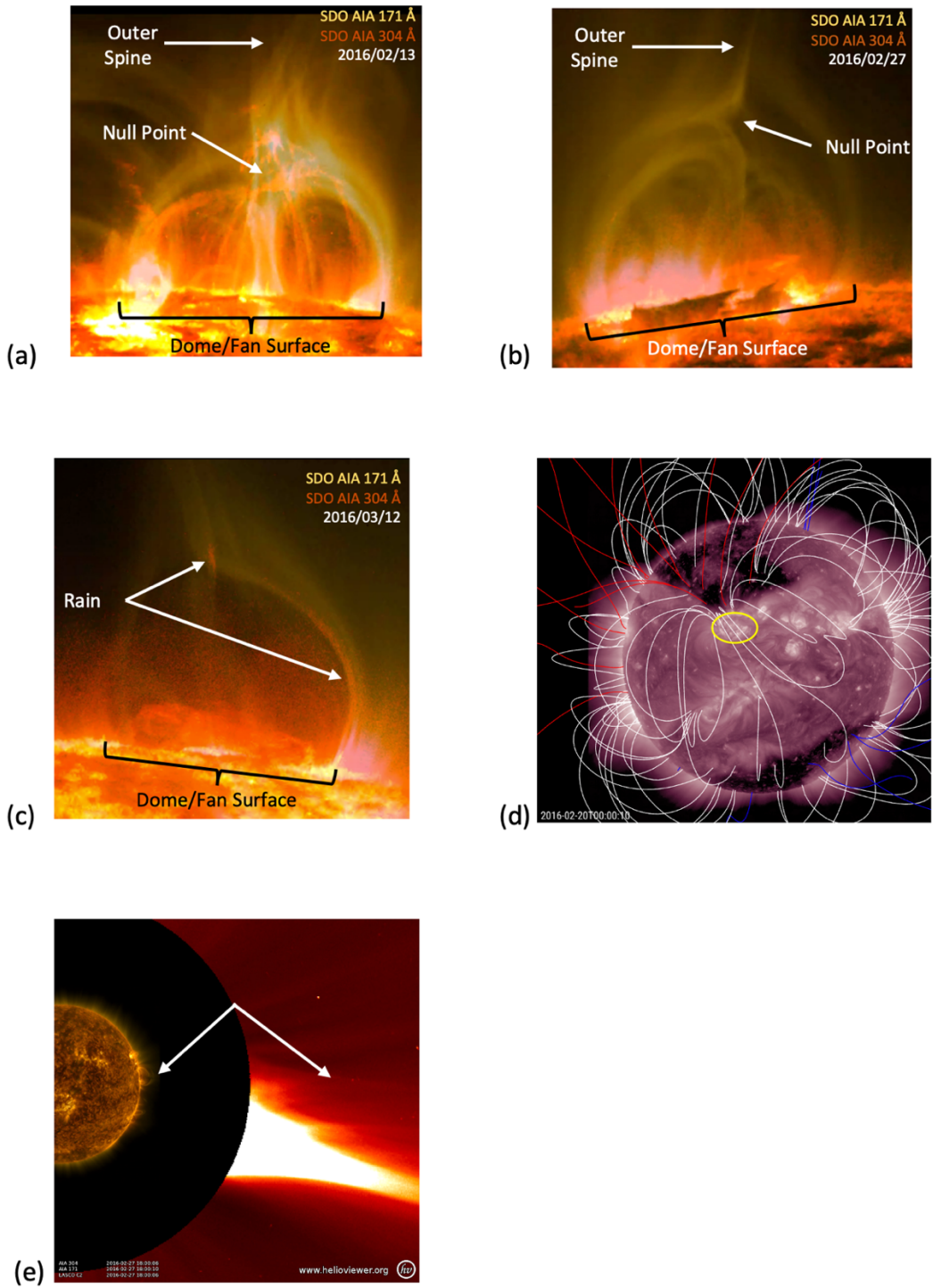

Figure 2: (a) Context image in SDO AIA 171 and 304 Å showing the early coronal signature of the RNPT on the eastern limb, previous to full formation of the underlying prominence. (b) Analogous image on the western limb, where the dark ring under the dome is the signature of the prominence, which has nearly finished forming at this point. (c) The RNPT once again on the eastern limb two weeks later, where the dome and spine are still intact but the prominence has already begun its slow rise phase; compare the height in this image to (b). (d) A PFSS extrapolation of the Sun with the RNPT circled in yellow; while it is placed on the coronal hole border easily visible in the AIA 211 Å context image, the extrapolation places the RNPT under the large equatorial helmet streamer. (e) Composite SDO AIA and SOHO LASCO C2 image with arrows showing the apparent spine of the RNPT extending out several solar radii above the surface. There are no other nearby structures which would produce the pseudostreamer-like spine in this image.

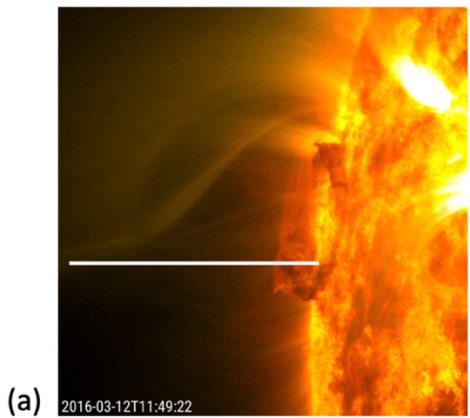

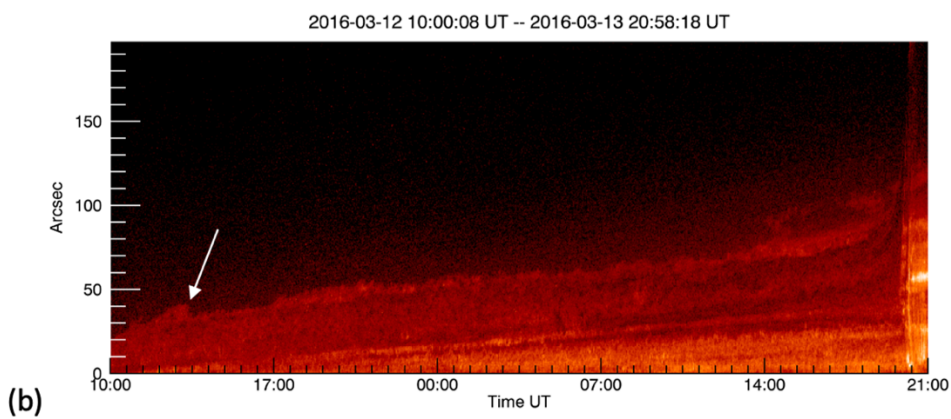

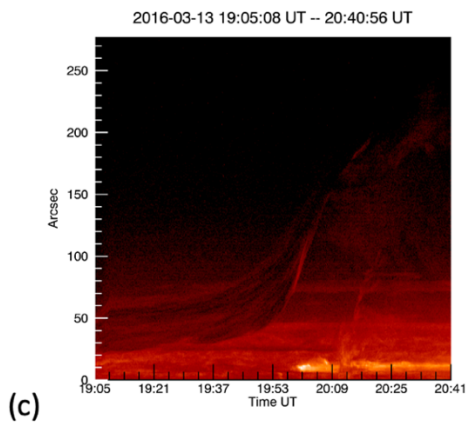
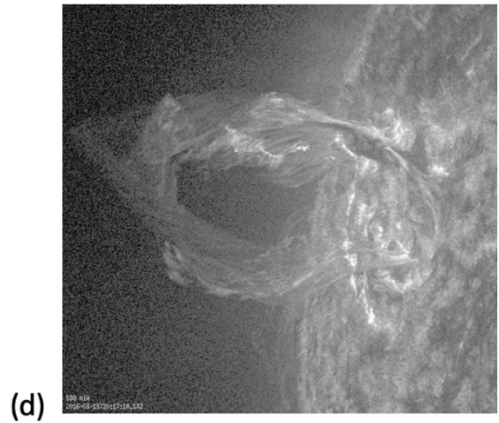

Figure 3: (a) Context SDO AIA 171 and 304 Å image showing the RNPT with a white line indicating the location of the cuts used to produce (b) and (c). (b) Time-distance plot showing predominantly the slow-rise phase of the prominence; solar rotation has been minimized in this stack of images, although not entirely removed. The residual motion was taken into account in velocity calculations. The white arrow indicates a draining event of filamentary material that rose more rapidly than the main prominence. (c) Time-distance plot showing the rapid-rise phase of the prominence. (d) Processed SDO AIA 304 Å data showing the extent of the bulk of the prominence material's rise; this links to an animation in the online version, showing that this material does not escape but drains back onto the western half of the prominence. The animation is 140 seconds in duration, and shows processed SDO AIA imagery of the entire rapid-rise eruption phase in the 171, 193, and 304 Å channels.

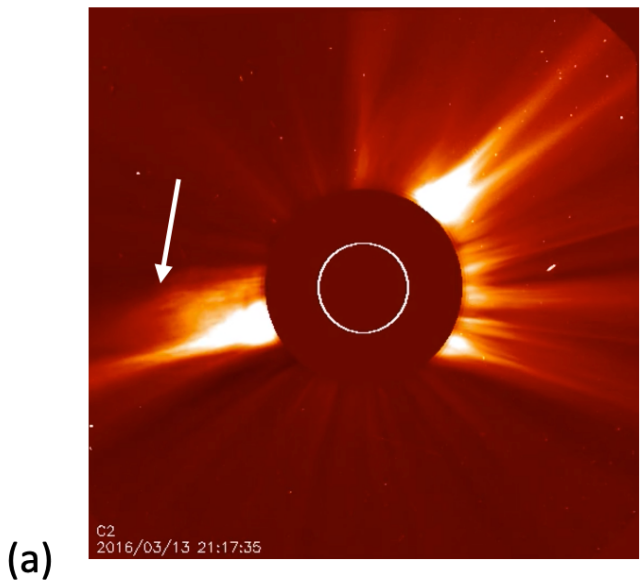 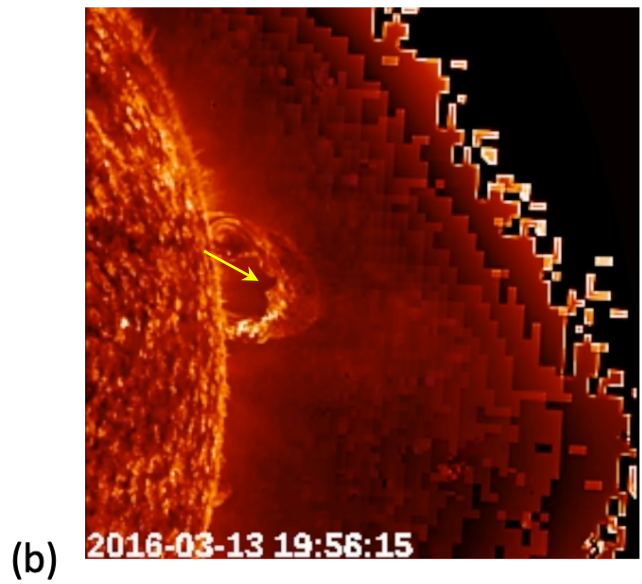

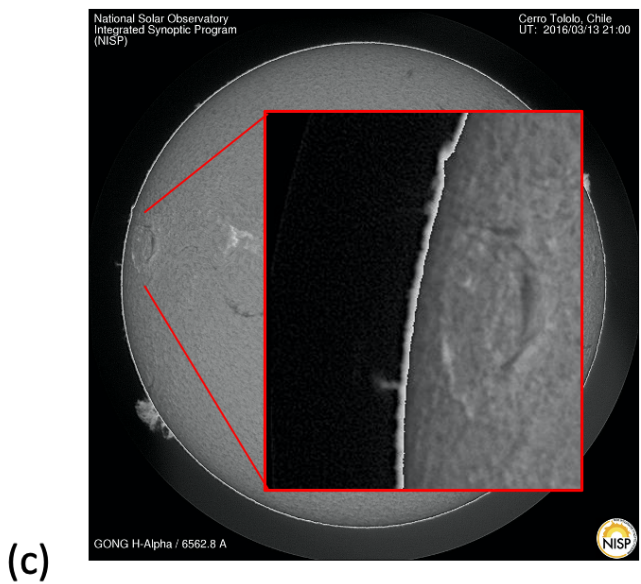 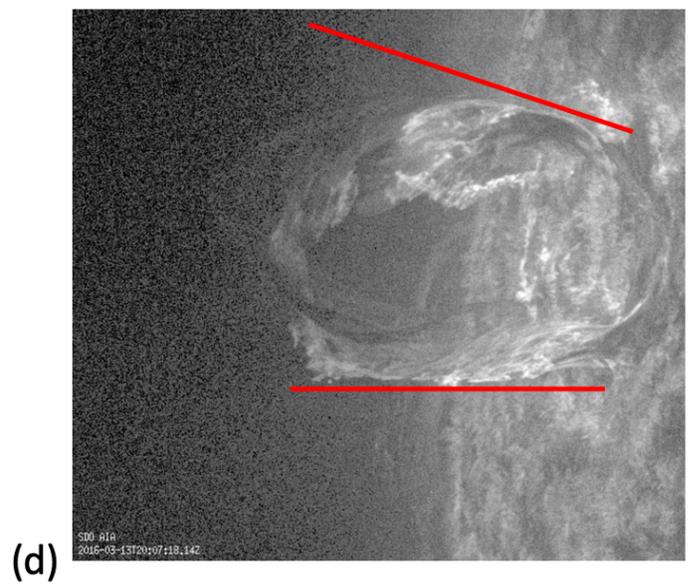

Figure 4: (a) SOHO LASCO C2 image showing the narrow CME ejected by the RNPT eruption; the white arrow indicates the CME. Note the close proximity of the eruption to the helmet streamer. (b) STEREO-A 304 Å image of the prominence eruption, showing reconnection along the southern loop leg analogous to that seen on the northern loop leg in the SDO AIA 304 Å imagery taken synchronously. The associated animation is 29 seconds in duration and shows the rapid-rise phase in STEREO-A EUVI imagery in the 195, 284, and 304 Å channels. (c) GONG H-alpha data from the Cerro Tololo site showing some of the prominence material near the surface, but no signal above it. Prominence material was visible in loops above the surface earlier, but had all drained back by this time and no material appeared that did not outline a closed loop. (d) Processed SDO AIA image showing the narrow bounds within which the eruption occurred; plasma in the eruption never exceeds the lines added in red, showing a high level of angular confinement.